\documentclass[twocolumn,english,prl,showpacs]{revtex4}
\usepackage[T1]{fontenc}
\usepackage[latin9]{inputenc}
\usepackage{float}
\usepackage{amsmath}
\usepackage{graphicx}
\usepackage{amssymb}
\usepackage{esint}
\usepackage{color}
\usepackage{babel}

\begin{document}

\title{Realization of $SU(N)$ Kondo effect in strong magnetic field}

\author{Assaf Carmi,$^{1, 2}$ Yuval Oreg,$^1$ and Micha Berkooz$^2$\\
\textit{$^1$Department of Condensed Matter Physics, Weizmann Institute
of Science, Rehovot, 76100, Israel\\$^2$Department of Particle Physics and Astrophysics, Weizmann Institute
of Science, Rehovot, 76100, Israel}}
\begin{abstract}
In this paper we suggest a realization of the $SU(N)$ Kondo
effect, using quantum dots at strong magnetic field. We propose using
edge states of the quantum Hall effect as pseudo spin that interact
with multiple quantum dots structures. In the suggested realization
one can access each pseudo spin separately and hence may perform a
set of experiments that were impossible until now. We focus on the
realization of $SU(2)$ and $SU(3)$ Kondo and find in the unitary limit a conductivity of 3/4
quantum conductance in the $SU(3)$ case.
\end{abstract}

\pacs{71.27.+a, 72.10.Fk, 75.20.Hr}

\maketitle

{\it Introduction:} At low temperatures, magnetic impurities inside a metallic host can significantly affect the behavior of the conduction electrons. This phenomenon, known as the Kondo effect (KE), has been studied for more than forty years \cite{Hewson}. In the last decade, realizations of KE with quantum dots (QDs) have enabled many new experiments, including measurements of transport properties such as conductance, phase shift and noise \cite{nature.391.156,science.281.540,PhysRevB.77.241303,PhysRevLett.100.226601}. KE in QDs can be achieved when a QD
is attached to leads. As we decrease the temperature to below the charging energy of the dot the conductance is suppressed due to Coulomb blockade \cite{0953-8984-16-16-R01}. If the ground state of the dot carries spin 1/2, lowering the temperature even further increases the conductance due to the KE, reaching $2e^2/h$ at zero temperature. The temperature at which the behavior of the conductance changes, denoted by $T_K$, is called the Kondo temperature. Weakly breaking the spin $SU(2)$ symmetry (e.g. with a small magnetic field, $B\ll T_K$) does not change the hierarchy of energy scales and the low energy nature of the Kondo physics \cite{0953-8984-16-16-R01}.

In the KE, a single channel of electrons, with spin up and spin down, interacts and screens a single level dot with spin up or spin down. This is the single channel $SU(2)$ KE. Here we generalize this system to have $N$ flavors instead of spin up and spin down such that we realize the $SU(N)$ KE \cite{Affleck,PhysRevB.80.125304,PhysRevB.80.155322}. This should not be confused with the situation of the k-channel $SU(2)$-KE \cite{refId}. There are therefore $N$ species of spin-less-like electrons in the leads, which are rotated into each other by $SU(N)$ transformations. This model is called the $SU(N)$-level 1 model, and it flows to one of $N$ possible renormalization group (RG) fixed points, depending on the details of the impurity. All these fixed points are described by Fermi-liquid (FL) theories. However, physical systems that exhibit $SU(N>2)$ KE are rare, usually complex and in practice limited to $N=4$, where the spin and another degree of freedom are exploited to form a four-fold degeneracy. Examples of such systems are: Double QD systems \cite{PhysRevLett.90.026602,PhysRevB.69.045326,PhysRevB.71.115312,PhysRevLett.93.017205},
triangular triple QD systems \cite{PhysRevB.80.155330}
and carbon nanotubes \cite{PhysRevLett.95.067204,nature.434.484,PhysRevB.74.205119}.

\begin{figure}
\begin{tabular}{cc}
\includegraphics[bb=100bp 235bp 470bp 645bp,clip,scale=0.28]{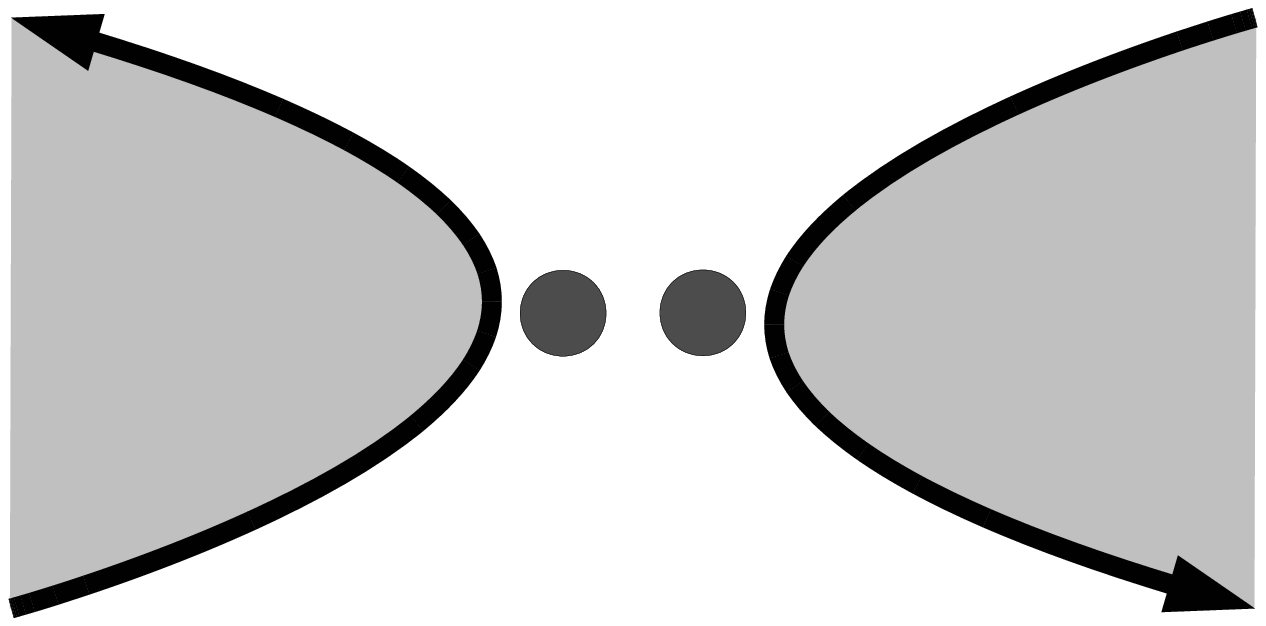} & \includegraphics[bb=60bp 235bp 430bp 645bp,clip,scale=0.28]{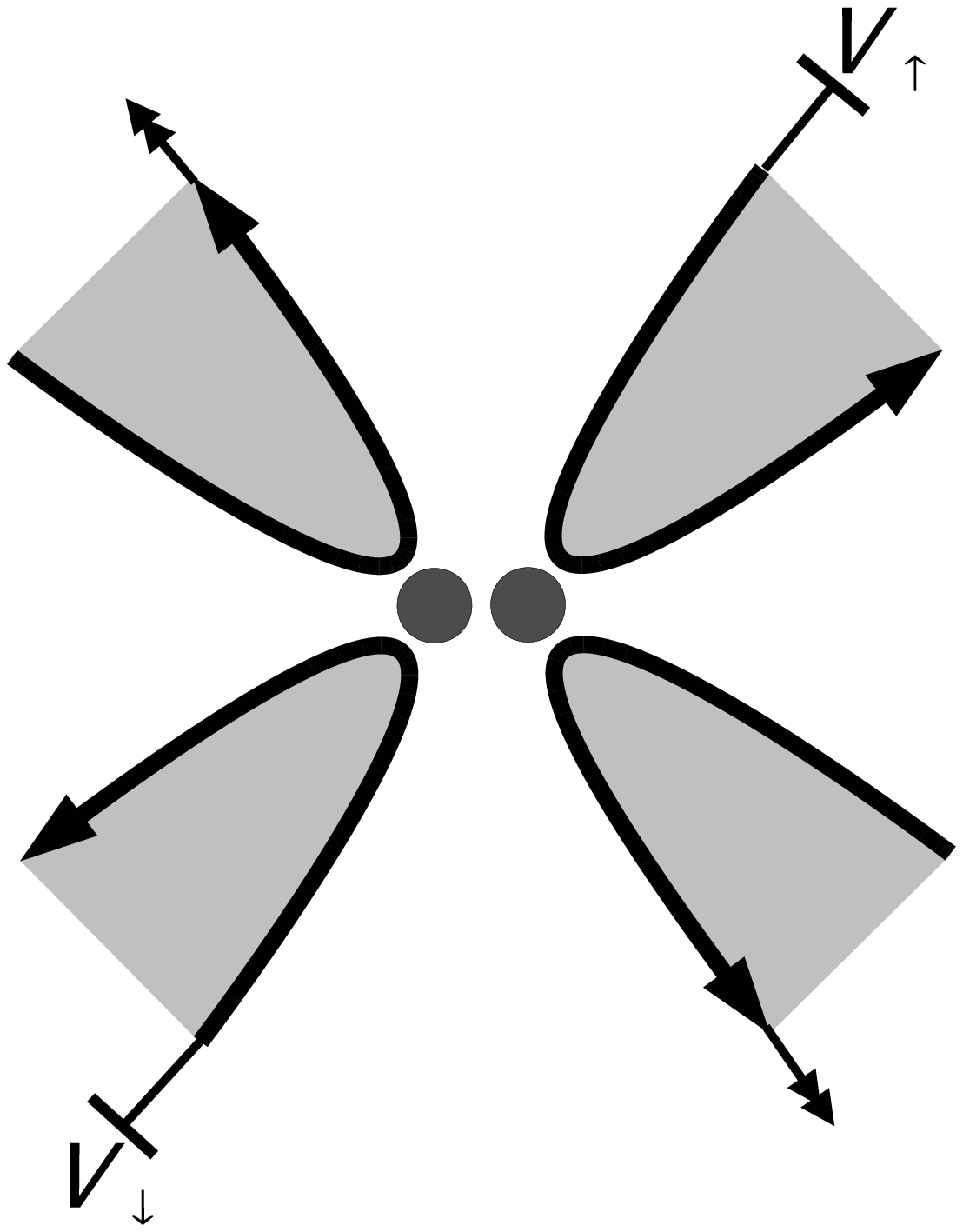}\tabularnewline
(a) & (b)\tabularnewline
\includegraphics[bb=40bp 260bp 468bp 625bp,clip,scale=0.28]{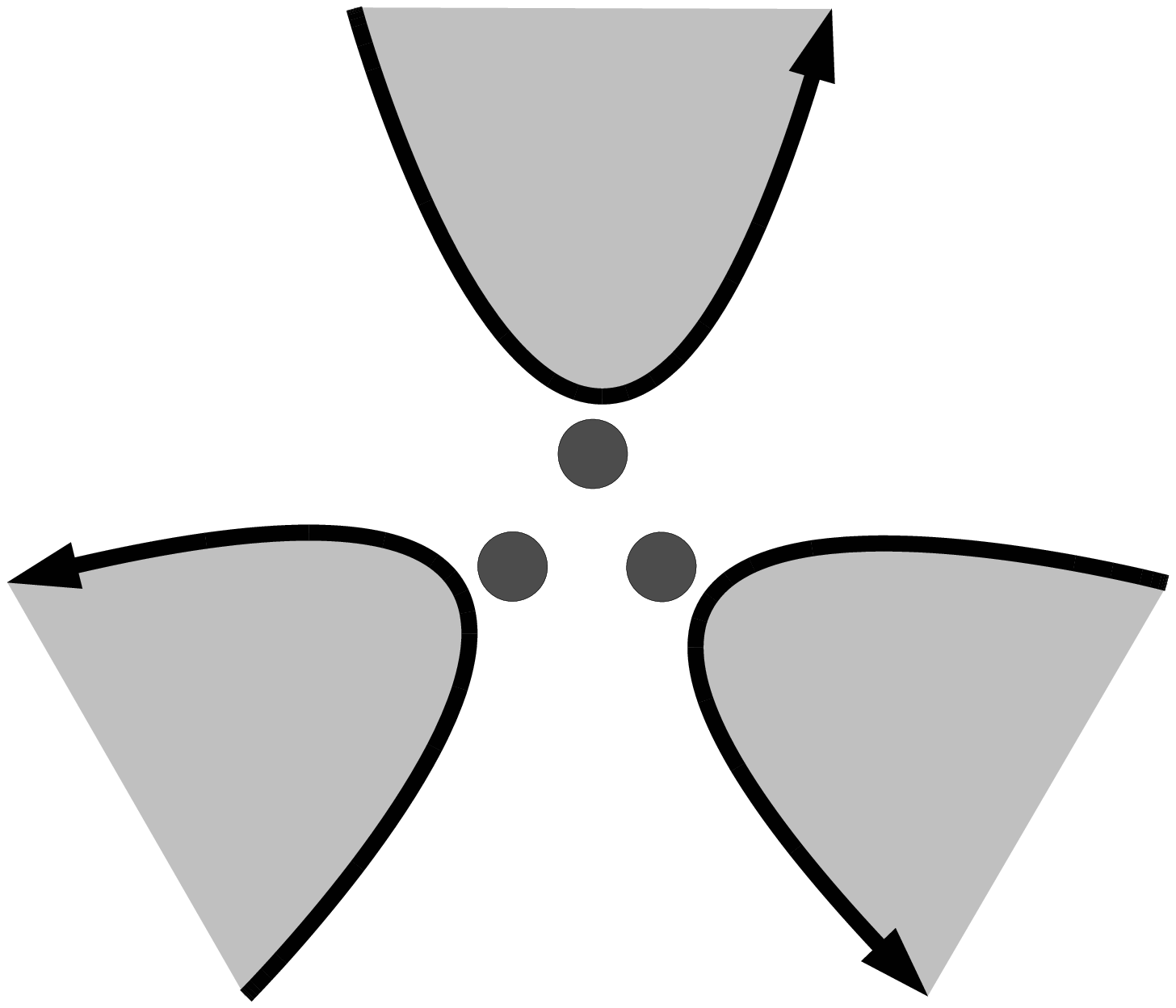} & \includegraphics[bb=40bp 260bp 468bp 625bp,clip,scale=0.28]{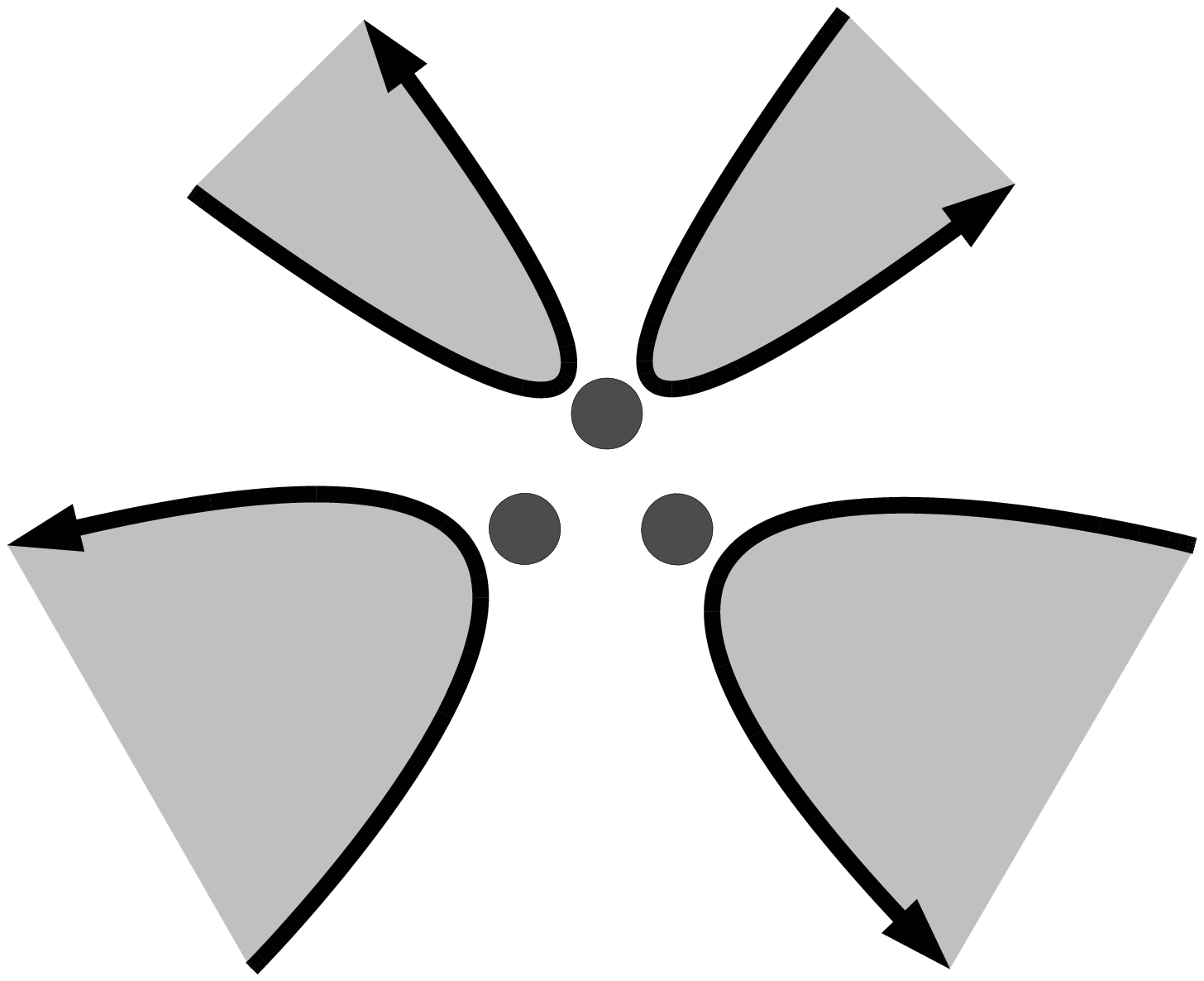}\tabularnewline
(c) & (d)\tabularnewline
\end{tabular}\caption{
\textbf{(a)} {\rm Realizing $SU(2)$ Kondo} using two edge states and
two quantum dots. Electrons can tunnel from each edge state to its nearest
dot. There is a strong coulomb interaction between the dots. The left/right sub-systems act as different
pseudo spins.
This realization {\it allows phase shift measurements}. \textbf{(b)} {\rm Realizing $SU(2)$ Kondo} using four edge states and two quantum dots. The edge states are split to {\it allow an independent transport measurement} of each pseudo-spin. \textbf{(c)} {\rm Realizing $SU(3)$ Kondo} using three edge states and three quantum dots with
strong coulomb interaction between them. Electrons can tunnel from
edge state to the nearest dot only. The three sub-systems act as different flavors. \textbf{(d)} {\rm
Realizing $SU(3)$ Kondo} using four edge states and three quantum
dots. Here, we split the upper edge state to {\it allow transport measurements}
between the two new upper edge states.\label{fig:Realizations}}
\end{figure}

In this paper we suggest a new realization of the KE and its generalization to $SU(N)$, using edge states of the quantum Hall effect that
interact with QDs (see Fig. \ref{fig:Realizations}). The basic building block that we use is a QD coupled to the edge state of an integer quantum Hall liquid.
The Hamiltonian of a single building block is therefore:
\begin{equation}
H_{1}=\sum_{k}\epsilon_{k}\psi_{k}^{\dagger}\psi_{k}+\epsilon_{g}d^{\dagger}d+(W\sum_{k}\psi_{k}^{\dagger}d+h.c.).
\end{equation}
The edge state is described by a chiral spinless fermion, $\psi$. The operator $d^{\dagger}$ populates the QD, which is assumed to have a single energy level
$\epsilon_{g}$ that is controlled by an outer gate. $W$ is the tunneling coefficient. Consider $N$ such building blocks as drawn in Figs. \ref{fig:Realizations}a,\ref{fig:Realizations}c for $N=2,3$. We assume a large repulsive interactions between the dots and there is no tunneling between them. In the $N=2$ case, the two blocks act as up and down pseudo-spins and the system realizes the $SU(2)$-Kondo physics. By using $N$ building blocks, we obtain the $SU(N)$-Kondo. For concreteness, in this manuscript we focus on the case $N=3$
as it is relatively easy to construct, yet rich enough to exhibit most of the features of the larger $N$ cases.

The systems that we suggest contain QDs with strong Coulomb interactions between them but very small tunneling between different dots (smaller than $T_K$). Double QDs systems fulfilling even stricter conditions already exist \cite{PhysRevB.74.233301,PhysRevLett.98.056801,hubel:102101,PhysRevLett.101.186804} with good control over the couplings to the leads, the gate voltages, and with practically zero tunneling between the dots. The implementation of KE due to the Coulomb interactions in those systems was demonstrated \cite{PhysRevB.74.233301,hubel:102101,PhysRevLett.101.186804}. Promising triple QDs systems are also experimentally available \cite{PhysRevB.76.075306,gaudreau:193101} making the current proposed realization of $SU(3)$-Kondo reasonably possible. We want to stress that obtaining the $SU(N)$ Kondo physics at low energies (IR) does not require a high accuracy of the symmetry between the $N$ different blocks. As long as the deviations from the symmetry are not relevant under RG, the system exhibits Kondo physics in the IR.

Besides the ability to realize $SU(N)$ Kondo, the systems that we suggest
have more advantages: First, as our realization is based on edge states
it can be easily integrated into an electronic Mach-Zehnder interferometer,
allowing accurate phase shift measurements. Second, our realization allows measurements
of a single pseudo-spin (or generally a single flavor) transport. Third, we can break the $SU(N)$ symmetry, potentially allowing a non Fermi liquid behavior such as in the two impurity Kondo model and its generalizations \cite{Affleck}. Fourth, it paves the way to possible generalizations to fractional quantum Hall edges, which may show a richer structure.

In the bulk of the manuscript we show the following:
\newline\textbf{-} In the first part we discuss the possible FL theories that can be obtained in the suggested $SU(N)$ systems and discuss the accuracy of the required symmetry.
\newline\textbf{-} In the second part we calculate the single flavor transport properties in the $N=2,3$ systems at relatively high energies (UV). We show that these are different from the total current transport properties. For example at the $SU(2)$ case, single pseudo-spin transport exhibits a Fano factor $F=7/9$, whereas the total current Fano factor is $F=5/9$ when both spins are equally biased.
\newline\textbf{-} In the third part we discuss the IR limit and show that the conductivity forms a Kondo-like plateaus with fractional conductivities $\frac{e^{2}}{h}\sin^2(\pi/N)$. For example for $N=3$ it gives $\frac{3}{4}\frac{e^{2}}{h}$. We also show that unlike the total flavor current, distinct flavor current-current cross-correlators receive $\mathcal{O}(1/T_K)$ corrections making them accessible to experiments at finite frequency.

{\it A new realization of} $SU(N)$: The Hamiltonian that describes a system
with $N$ building blocks is:
\begin{eqnarray}
H_{N} & = & \sum_{\alpha=1...N}\sum_{k}\epsilon_{k}\psi_{k\alpha}^{\dagger}\psi_{k\alpha}+\epsilon_{g}\sum_{\alpha}d_{\alpha}^{\dagger}d_{\alpha}\label{eq:H_n}\\
 & + & \left(W\sum_{\alpha}\sum_{k}\psi_{k\alpha}^{\dagger}d_{\alpha}+h.c.\right)+\sum_{\alpha<\beta}U_{\alpha\beta}d_{\alpha}^{\dagger}d_{\alpha}d_{\beta}^{\dagger}d_{\beta}.\nonumber
\end{eqnarray}
The indices $\alpha,\beta$ label the different building blocks and $U_{\alpha\beta}$ is the charging energy between the dots $\alpha$ and $\beta$. We have made the assumption of identical sub-systems (that is $\epsilon_k$ ,$W$ and $\epsilon_g$ are flavor-independent), and we assumed that there is no tunneling between the dots.

These assumptions are not crucial. Small differences between the energies of the dots and small changes in the tunneling coefficients have an effect similar to a small magnetic field and a small exchange field in the familiar $SU(2)$ KE. We therefore only require that the differences are small compare to the Kondo temperature: $\Delta\epsilon_g\ll T_K$ and $|\Delta W|^2/\epsilon_g\ll T_K$. Jumping ahead, these deviations from the symmetry will at most change the IR behavior by marginal operators and the conductance through flavor $i$ would be given by $\frac{e^2}{h}\sin^2(\pi n_i)$ ($n_i$ is the average occupation of the $i$th dot). Small tunneling terms between the dots ($t_{\alpha\beta}d^{\dagger}_\alpha d_\beta+h.c.$) can be mapped to small differences in the energies of the dots by simply rotating the dots' basis, we only require that $|t_{\alpha\beta}|\ll T_K$.

Let us assume temporarily identical charging energies, $U_{\alpha\beta}=U$. In this case the Hamiltonian \eqref{eq:H_n} possesses a $SU(N)$ symmetry acting on the states of the dots by $S^{i}=\sum_{\alpha,\beta}d_{\alpha}^{\dagger}T_{\alpha\beta}^{i}d_{\beta}$, where $T^i$ are the $N\times N$ generators of $SU(N)$.
Having $m$ electrons in the dots amounts to an impurity in a representation of $SU(N)$ with $m$ anti symmetrized indices, each in the fundamental representation \cite{PhysRevB.80.125304,Affleck}. We will discuss below to what extent different $m$'s can be realized, but for now we consider a general $m$.
In this case, after a Schrieffer-Wolff transformation the Hamiltonian \eqref{eq:H_n} is:
\begin{equation*}
H_{K}=\sum_{\alpha}\sum_{k}\epsilon_{k}\psi_{k\alpha}^{\dagger}\psi_{k\alpha}+ J\sum_{i}\sum_{\alpha,\beta}\sum_{k,k'}\psi_{k\alpha}^{\dagger}T_{\alpha\beta}^{i}\psi_{k'\beta}S^{i},
\end{equation*}
where $J\sim W^{2}/U$ and the index $i$ runs over the $N^2-1$ $SU(N)$ generators.
Conformal field theories with $SU(N)$ Kac-Moody level 1, which is the case here, have $N$ gaussian invariant boundary states that correspond to $N$ possible FL fixed points of the RG. The natural conjecture, supported by the Friedel sum rule, is that populating the array of dots by $0\leq m<N$ electrons leads to a flow to each of these boundary states, characterized by boundary conditions $\psi_i(x=0^+)=e^{2i\frac{m\pi}{N}}\psi_i(x=0^-)$.

We now discuss the feasibility of the $SU(N)$ picture and the possible $m$'s where we release the restriction of identical charging energies. Here we describe only the limit $|\epsilon_g|\ll U_{\alpha\beta}$, which is similar to the asymmetric Anderson model \cite{PhysRevB.80.155322} of the $SU(2)$ case. In the supplementary material we give a more complete description with no restriction on the ratio $|\epsilon_g|/U_{\alpha\beta}$. In the exact $U_{\alpha\beta}\rightarrow\infty$ limit, the total number of electrons in the dots is limited to zero or one, and the system (\ref{eq:H_n}) has a $SU(N)$ symmetry. The reason is that only states with higher occupations are sensitive to the $SU(N)$ breaking by the different $U_{\alpha\beta}$'s but these occupations are never reached. The system therefore flows to a $SU(N)$ symmetric fixed point. In the physical finite $U_{\alpha\beta}\gg \epsilon_g$ case, the $SU(N)$ symmetry breaking operators are marginal at most: $\sum_{\alpha\beta}\lambda_{\alpha\beta}\psi_{\alpha}^{\dagger}\psi_{\beta}$ with $\lambda_{\alpha\beta}\sim\epsilon_g/U_{\alpha\beta}$. In the supplementary material we show that under a mild restriction on the geometry of the dots we flow to a $SU(N)$ symmetric fixed point in the IR with no limitations on $|\epsilon_g|/U_{\alpha\beta}$. We have showed the feasibility of the the $m=1$ case, similar arguments can be used to show the feasibility of the particle-hole symmetric $m=N-1$ case.

\begin{figure}
\begin{tabular}{cc}
\includegraphics[bb=115bp 506bp 1000bp 1177bp,clip,scale=0.13]{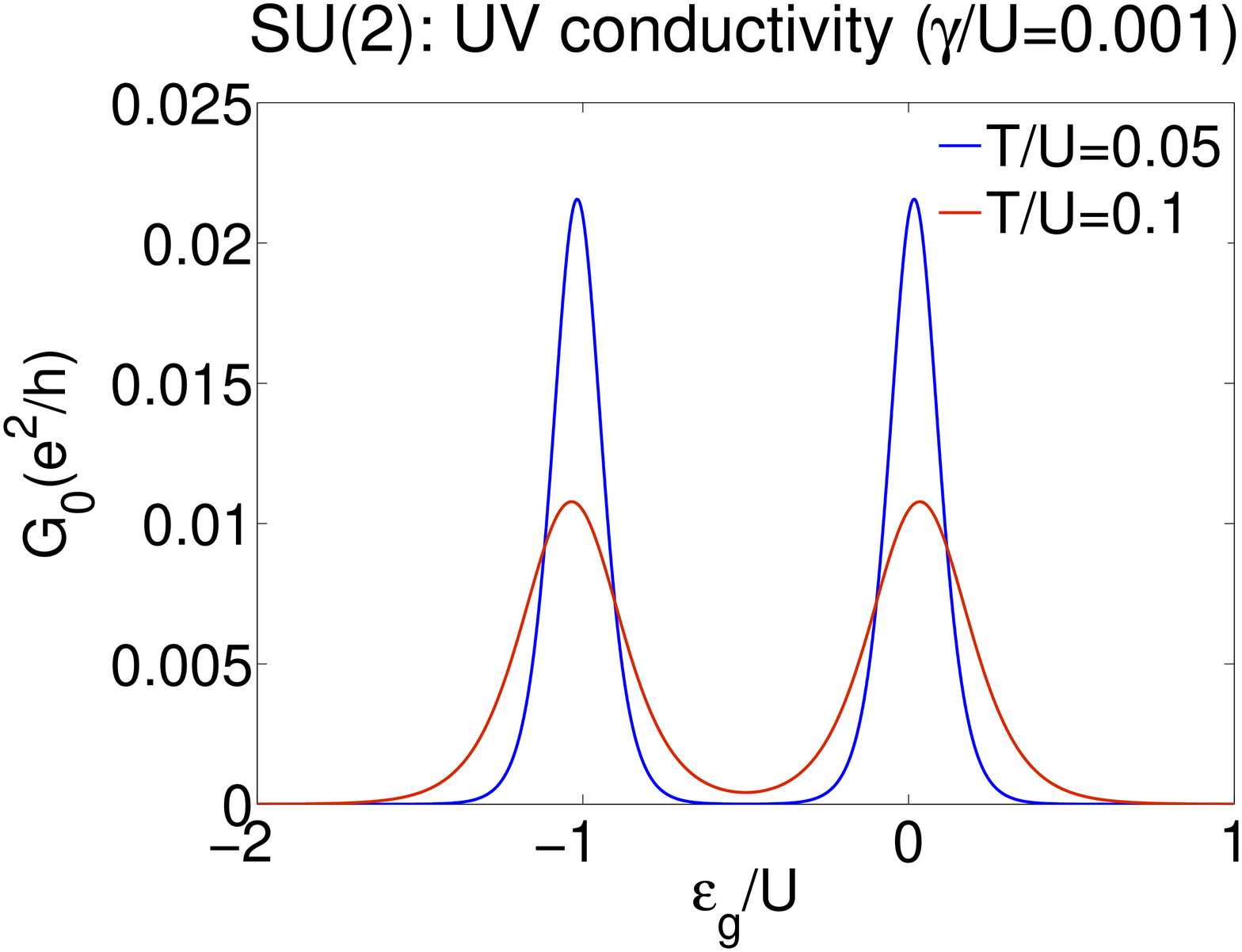} & \includegraphics[bb=115bp 506bp 1000bp 1177bp,clip,scale=0.13]{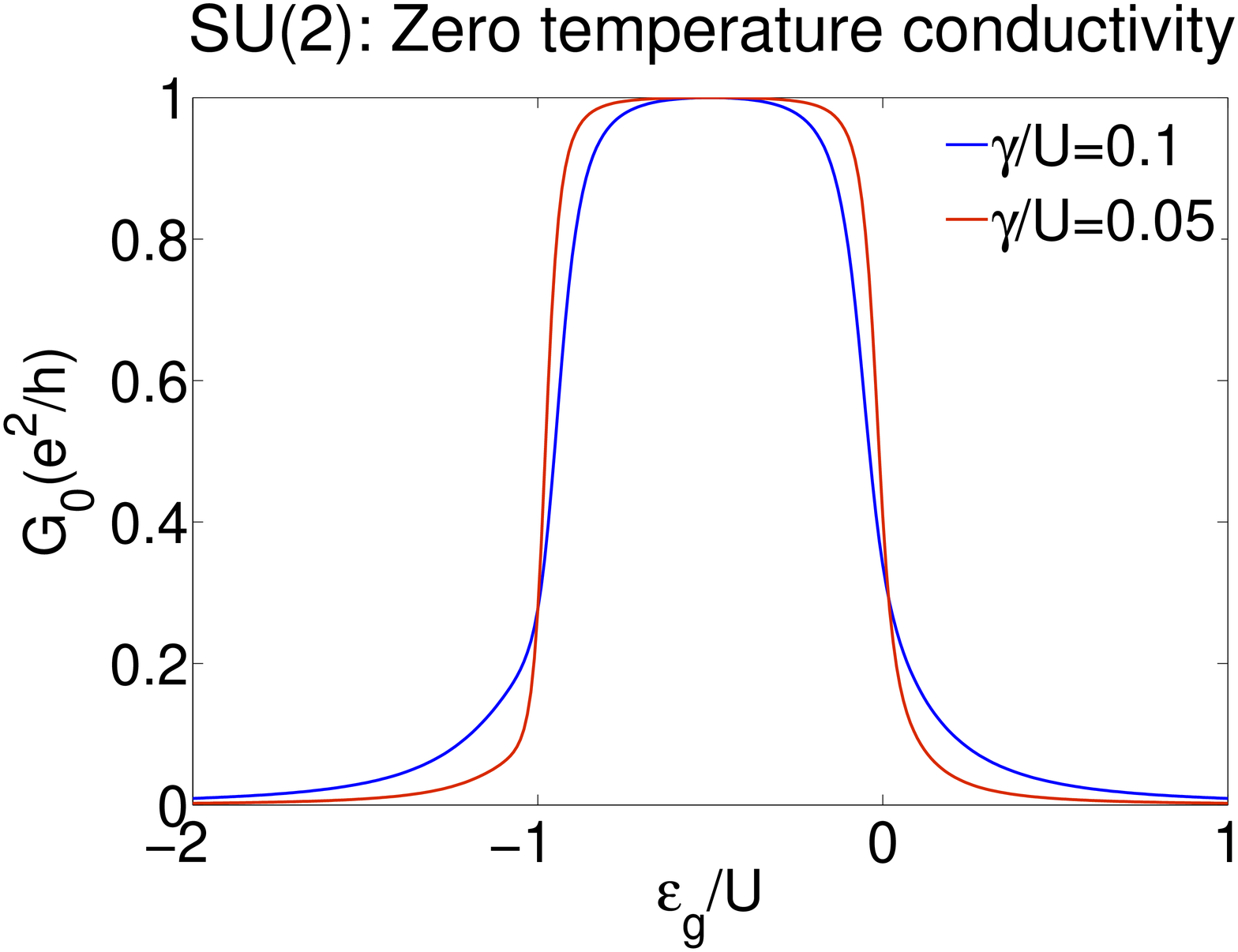}\tabularnewline
(a) & (b)\tabularnewline
\includegraphics[bb=115bp 506bp 1000bp 1177bp,clip,scale=0.13]{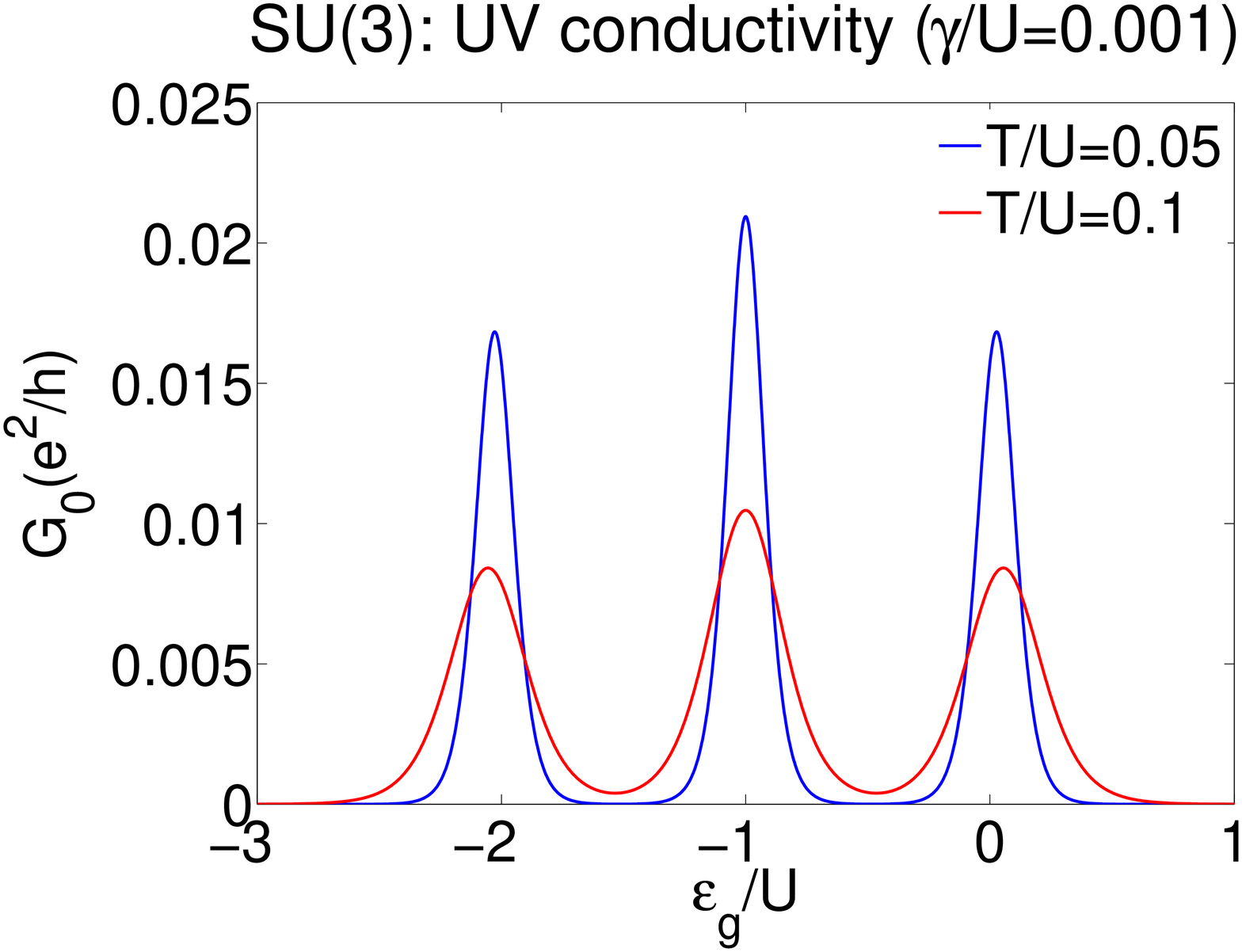} & \includegraphics[bb=115bp 506bp 1000bp 1177bp,clip,scale=0.13]{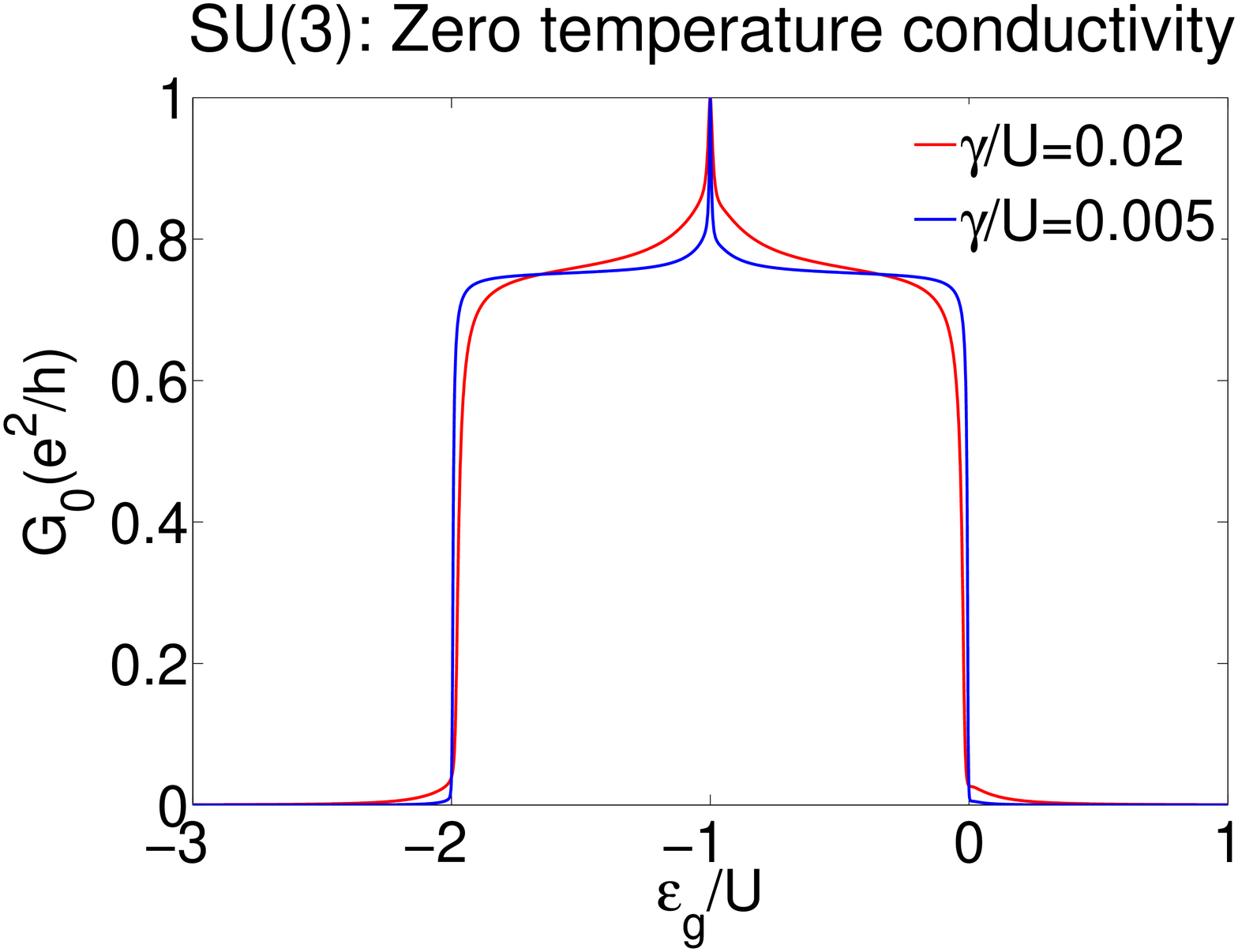}\tabularnewline
(c) & (d)\tabularnewline
\end{tabular}\caption{$SU(2)$ {\it system}: \textbf{(a)} The conductivity of a single
pseudo spin at high temperature ($T\gg\gamma$) for two different $T/U$ ratios. We choose the Fermi
energies of the leads to be zero. The Coulomb peaks are not exactly centered
around $0,\, -U$ due to the asymmetry between empty/occupied states
near the peaks. \textbf{(b)} Self consistent Hartree solution of the
zero temperature conductivity of a single pseudo spin for two different $\gamma/U$ ratios. $SU(3)$
{\it system}: \textbf{(c)} The conductivity of a single flavor at high
temperature ($T\gg\gamma$) for two different $T/U$ ratios choosing the Fermi energy to be zero.
The outer peaks are not exactly centered around $0,\,-2U$ due to
the asymmetry between empty/occupied states near the peaks. \textbf{(d)}
Self consistent Hartree solution of the zero temperature conductivity
of a single flavor for two different $\gamma/U$ ratios.\label{fig:conductances}}
\end{figure}

{\it Results}: As double QD systems are available and triple QD systems were also fabricated \cite{PhysRevB.76.075306,gaudreau:193101} we focus on systems
that realize $SU(2)$ and $SU(3)$ Kondo models. We first look at the UV
limit (high temperature or bias voltage) and later discuss the Kondo physics at the IR. Special attention
is given to the measurement of each pseudo-spin/flavor
independently. In order to measure transport properties in one flavor we split an edge into two different edges which are coupled to the same dot, as drawn in Fig. \ref{fig:Realizations}d. In equilibrium, one linear combination of the two edge states
is decoupled from the upper dot and the orthogonal linear combination plays the role of a single edge state that is coupled to the upper dot. Transport is measured by applying a voltage difference between the two edges allowing measurements of average current $\langle I\rangle$, zero frequency shot noise $S_{\rm shot}$, and the Fano factor $F=\frac{S_{\rm shot}}{2e\langle I\rangle}$ of each flavor separately. Measuring correlations between two flavors can be carried out by splitting two edges as drawn in Fig. \ref{fig:Realizations}b, where we can apply different bias voltages on different flavors.

{\it UV description}: We assume a constant density of states $\nu$ in the edges and a very weak coupling between the dots and the edge states ($\nu|W|^2\ll T,$
or $\nu|W|^2\ll V_{\mathrm{bias}}$), and use the rate equations method
\cite{PhysRevB.49.10381} to calculate the conductivity and the current
noise.

Figs. \ref{fig:conductances}a, \ref{fig:conductances}c plot
the Coulomb peaks structures of the $SU(2)$ and $SU(3)$ systems. Contrary to naive expectations, at finite temperature the outer peaks are not centered around $\epsilon_{F},$
$\epsilon_{F}-U$, in the N=2 case, or around $\epsilon_{F},$ $\epsilon_{F}-2U$
in the N=3 case. Due to the asymmetry between occupied and empty states near the outer peaks, the peaks are shifted by $T\ln(N)/2$. We discuss this point in the supplementary material.

The Fano factor of a single spinful QD which
is attached to two spinful leads is $5/9$ at the Coulomb peaks. In the system of Fig. \ref{fig:Realizations}b it corresponds to applying the same bias voltage on the two pseudo spins and measuring the total current ($I_{\uparrow}+I_{\downarrow}$). However, in the systems of Figs. \ref{fig:Realizations}b and \ref{fig:Realizations}d, we can apply distinct bias voltages and measure distinct currents for the different pseudo spins or flavors. The Fano factors that we obtain are:\[
F_{2}=7/9,\,\,\,\,\, F_{3}^{o}=7/8,\,\,\,\,\, F_{3}^{c}=11/18,\]
where $F_{2}$ is the Fano factor near the N=2 Coulomb peaks, and
$F_{3}^{o}$ ($F_{3}^{c}$) is the Fano factor near a outer (central)
peak of the N=3 case.
The Fano factor is related
to the effective charge of the current pulses that cross the dots,
it is therefore counter-intuitive to find different Fano factors in the total and single flavor currents. The difference is a result of the zero frequency cross-correlations between the different flavors. Using rate equations we find in the $N=2$ case: \begin{eqnarray}
S_{\uparrow\downarrow}(0) & \equiv & \lim_{\omega\rightarrow0}\int dte^{i\omega t}2\left(\left\langle I_{\uparrow}(t)I_{\downarrow}(0)\right\rangle -\left\langle I_{\uparrow}\right\rangle \left\langle I_{\downarrow}\right\rangle \right)\\
 & = & -\frac{2\pi\nu|W|^2 e^{2}}{\hbar}\frac{4}{27}\tanh\frac{eV_{\uparrow}}{4T}\tanh\frac{eV_{\downarrow}}{4T},\nonumber \end{eqnarray}
where $V_{\uparrow,\downarrow}$ are the pseudo-spin bias voltages.
Similarly, in the N=3 case, for $\alpha\neq\beta$:
\begin{equation}
S_{\alpha\beta}^{(i)}(0)=-\frac{2\pi\nu|W|^2 e^{2}}{\hbar}R^{(i)}\tanh\frac{eV_{\alpha}}{4T}\tanh\frac{eV_{\beta}}{4T},
\end{equation}
where the index $i=o,c$ labels a central (outer) peak, and $R^{o}=1/16,$
$R^{c}=1/27.$

{\it The IR Kondo fixed point:} In Figs. \ref{fig:conductances}b,\ref{fig:conductances}d we
depict the zero temperature conductance of a single pseudo spin/flavor in
the $N=2,3$ cases. The conductance
was calculated using self consistent Hartree approximation. We average
over the solutions of the self consistent equations for the average
occupations of the levels \cite{PhysRevB.72.125316}:
$n_{\alpha}=\frac{1}{2}-\frac{1}{\pi}\tan^{-1}\frac{E_{d}+U(\sum_{\beta\ne\alpha}n_{\beta})}{\gamma}$, where $\gamma$ represents the identical width of the levels. The Friedel sum rule and spin/flavor symmetry are then used to determine the phase shift: $\delta_{\alpha}=\delta=\pi\sum_{\alpha=1}^{N}n_{\alpha}/N$ and the conductance $\sin^{2}(\delta)$. Note that in our setup this phase shift can be measured directly in a Mach-Zehnder interferometer.

Lets focus on the N=3 case (Fig. \ref{fig:conductances}d): When the
gate voltage is tuned to have only a single electron or a single
hole in the three dots, the conductance is enhanced due to the flavor-interaction
and the Coulomb valleys disappear. An unusual plateau is created instead
of the three peaks of the UV limit. As long as the dots are occupied
by a single electron or a single hole we find the conductivity:
\begin{equation}
G_{SU(3)\mathrm{-Kondo}}=\frac{3}{4}\frac{e^{2}}{h}.
\end{equation}
At the exact particle hole symmetric point, $\epsilon_{g}=\epsilon_{F}-U$,
we find a full $e^{2}/h$ conductivity and a sharp peak is formed,
indicating the larger symmetry.

{\it Edge states cross-correlations in the Kondo limit}: Finally, we will discuss the finite frequency flavors correlations in the IR limit. The cross-flavors correlations is a new observable made available by our realization, whereas it is not experimentally accessible in most existing realizations of KE. This cross-flavors correlations is of order $\omega/ T_K$ relative to the familiar large thermal noise in the total flavor. For $\omega=100$MHz and $T_K\sim1$K, $\omega/T_k\sim 1\%$ which, being the only contribution in the cross channel, can be readily observed.

Formally, we define $J_{i}(x)=\colon\psi_{i}^{\dagger}(x)\psi_{i}(x)\colon$ and:\[
J_{c}(x)=\sum_{i=1}^{N}J_{i}(x),\,\,\,\vec{J}_{(N)}(x)=\sum_{ij=1}^{N}\colon\psi_{i}^{\dagger}(x)\vec{T}_{ij}^{(N)}\psi_{j}(x)\colon,\]
where $N=2,3$, $\vec{T}^{(2)}$ are the three Pauli matrices and
$\vec{T}^{(3)}$ are the eight Gell-Mann matrices. The effective Hamiltonians
of the two systems are \cite{Affleck,Nosieres,PhysRevB.80.125304}:\begin{eqnarray}
H_{2}=\frac{1}{8\pi}\int dx\left(J_{c}^{2}(x)+\frac{1}{3}\vec{J}_{(2)}^{2}(x)+\lambda_{2}\vec{J}_{(2)}^{2}(x)\delta(x)\right),&&\\
H_{3}=\frac{1}{8\pi}\int dx\left(J_{c}^{2}(x)+\frac{3}{8}\vec{J}_{(3)}^{2}(x)+\lambda_{3}\vec{J}_{(3)}^{2}(x)\delta(x)\right),&&\end{eqnarray}
with $\lambda_{N}\sim1/T_{k}$. We calculate the noise functions:
\begin{equation}
S_{ij}(\omega;x,x')=\int e^{i\omega t}\left\langle \left\{ J_{i}(x,t),J_{j}(x',t')\right\} \right\rangle .\label{eq:Definition of noise}
\end{equation}
The noise functions receive $\mathcal{O}\left(\omega/T_{k}\right)$
correction ($i\neq j$):\begin{eqnarray}
S_{ii}^{(N)}(\omega;x,x') & = & e^{-i\omega(x-x')}\frac{\omega}{\pi}\coth\frac{\beta\omega}{2}\left(1-i\widetilde{\lambda}_{N}\frac{\omega}{\pi}F_{xx'}\right),\nonumber \\
S_{ij}^{(N)}(\omega;x,x') & = & \frac{i\widetilde{\lambda}_{N}}{N-1}e^{-i\omega(x-x')}\left(\frac{\omega}{\pi}\right)^{2}\coth\frac{\beta\omega}{2}F_{xx'},\end{eqnarray}
with $F_{xx'}=\theta(x)-\theta(x')$, $N=2,3$, $\widetilde{\lambda}_{2}=6\lambda_{2}$ and $\widetilde{\lambda}_{3}=\frac{32}{3}\lambda_{3}$.
Due to $F_{xx'}$, $S_{ij}$ receive correction only if the two currents are measured at two different sides of the dots.

{\it Acknowledgement} We would like to thank Z. Ringel, N. Lezmy and E. Sela for useful discussions. This work is supported in part by the BSF, DIP, ISF and Einstein-Minerva center.

\section{Supplementary material}
\subsection{The feasibility of the $SU(N)$ picture with finite charging energies}
We discuss the feasibility of the $SU(N)$ picture, here we do not require that $\epsilon_g\ll U_{\alpha\beta}$. We make a distinction between two cases:

$SU(N\leq3)$ Kondo: We arrange the dots on the vertices
of a triangle, such that the distances between pairs of dots are roughly the same. We do not require that the triangle be exactly equilateral. However, it is convenient to start from this idealized case, and then verify that the dynamics of the system does not change significantly in the non strictly equilateral physical case. In the exactly equilateral case $U_{\alpha\beta}=U$, the Hamiltonian:
\begin{eqnarray*}
H_{N} & = & \sum_{\alpha=1...N}\sum_{k}\epsilon_{k}\psi_{k\alpha}^{\dagger}\psi_{k\alpha}+\epsilon_{g}\sum_{\alpha}d_{\alpha}^{\dagger}d_{\alpha}\label{eq:H_n_supp}\\
 & + & \left(W\sum_{\alpha}\sum_{k}\psi_{k\alpha}^{\dagger}d_{\alpha}+h.c.\right)+\sum_{\alpha<\beta}U_{\alpha\beta}d_{\alpha}^{\dagger}d_{\alpha}d_{\beta}^{\dagger}d_{\beta},
\end{eqnarray*}
is $SU(N)\times U(1)$ symmetric and the flow proceeds as we discussed in the manuscript. Small deviations from the point $U_{\alpha\beta}=U$ can then be analyzed using the IR theory where they are at most marginal.

$SU(N\geq4)$ Kondo: Unlike the $N\leq3$ case, we can no longer assume $U_{\alpha\beta}=U$ because the distances between the various pairs of dots, adjacent and non adjacent, cannot be made approximately equal. We can however arrange the dots on a two dimensional regular polygon (RP), then $U_{\alpha\beta}=U_{\alpha-\beta}$. In this case, the system has a $Z_N$ symmetry (rotation by $2\pi/N$) and a $U(1)^N$ symmetry associated with charge conservation of each flavor. In the next paragraph we show that the $Z_N\times U(1)^N$ symmetry ensures that the $SU(N)$ symmetric single occupation fixed point ($m=1$ or $m=N-1$) is attractive. In the physical case, the $Z_N$ symmetry is not exact, nevertheless analyzing the IR fixed point shows that symmetry breaking operators are marginal at most \cite{footnote_on_SU4}.

The stability of the fixed point in the $Z_N\times U(1)^N$ symmetric case proceeds as follows: The general marginal operator near the IR is of the form $\sum_{\alpha\beta}\lambda_{\alpha\beta}\psi_{\alpha}^{\dagger}\psi_{\beta}$. The $U(1)$ symmetry on each of the edges, i.e. charge conservation in each subsystem, implies invariance under independent multiplication of each $\psi_\alpha$ by a phase. Hence the matric $\lambda_{\alpha\beta}=\lambda_\alpha \delta_{\alpha\beta}$. The $Z_{N}$
symmetry then enforces that all the $\lambda_\alpha$ are equal. The only remaining permissible operator, $\lambda\sum_{\alpha}\psi_\alpha^{\dagger}\psi_\alpha$,
does not break the $SU(N)$ symmetry. Since the $SU(N)$ fixed point is attractive, there is a finite range of UV couplings in which the theory flows to this IR fixed point. In the physical case the $Z_N\times U(1)^N$ symmetry is approximate and the only exact symmetry is total charge conservation which is the sum of the $N$ different $U(1)$s. This symmetry in enough to exclude all the relevant operators in the theory ($\psi$ and $\psi^{\dagger}$). The allowed operators $\psi_\alpha^{\dagger}\psi_\beta$ are marginal at most.

This picture explains the experimental results in \cite{PhysRevLett.93.017205} and \cite{nature.434.484} where indications of $SU(4)$ Kondo were obtained starting from UV systems which lack the exact symmetry. The picture is further supported by a perturbative scaling analysis and by NRG computations for $N=4$ \cite{PhysRevLett.90.026602}, suggesting a large basin of attraction for this flow.

\subsection{Coulomb peaks structure of the conductivity at finite temperature}

Figs. 2a, 2c  of the manuscript plot
the Coulomb peaks structures of the conductivity of the $SU(2)$ and $SU(3)$ systems. Contrary to naive expectations, at finite temperature the outer peaks are not centered around $\epsilon_{F},$
$\epsilon_{F}-U$, in the N=2 case, or around $\epsilon_{F},$ $\epsilon_{F}-2U$
in the N=3 case ($\epsilon_{F}=0$ in the plots). Focusing
on the right most peak, the conductivity through a dot $\alpha$ is proportional to:
\begin{equation}
G_\alpha\sim(P_{0}(\epsilon_g)+P_{\alpha}(\epsilon_g))f(\epsilon_g)(1-f(\epsilon_g)),
\end{equation}
where $P_{0}=\left(1+Ne^{-\epsilon_{g}/T}\right)^{-1}$ is the probability of finding all the dots empty, $P_{\alpha}=(1-P_0)/N$ is the probability of finding an electron in the dot $\alpha$ (for $T\ll U$) and $f(\epsilon)$ is the Fermi Dirac distribution function of electrons in the edges. The term $(P_{0}(\epsilon_g)+P_{\alpha}(\epsilon_g))$ does not have a particle hole symmetry at finite temperature, therefore the peak is shifted from $\epsilon_{F}$, in this case by $T\ln(N)/2$. In
the $SU(3)$ system the point $\epsilon_g=\epsilon_{F}-U$ is a half-filling
point; the three dots are occupied on average by $3/2$ electrons.
Due to particle hole symmetry, the Coulomb peaks structure is
symmetric around $\epsilon_g=\epsilon_{F}-U$ and the central peak is pinned at this half-filling point.

\bibliographystyle{apsrev}
\bibliography{paper_Realization_of_Kondo}

\begin{thebibliography}{28}
\bibitem[{foo()}]{footnote_on_SU4}
\bibinfo{note}{Notice that if the system is tuned to have $1<m<N-1$ electrons
  in the dots then without further tuning the ground states do not form a
  representation of $SU(N)$ and therefore the theory is not $SU(N)\times U(1)$
  symmetric.}
\end{thebibliography}

\end{document}